\documentstyle[12pt,axodraw]{article} 
\catcode`@=11
\newcount\@tempcntc
\def\@citex[#1]#2{\if@filesw\immediate\write\@auxout{\string\citation{#2}}\fi
\@tempcnta\z@\@tempcntb\m@ne\def\@citea{}\@cite{\@for\@citeb:=#2\do
{\@ifundefined
{b@\@citeb}{\@citeo\@tempcntb\m@ne\@citea\def\@citea{,}{\bf
?}\@warning
       {Citation `\@citeb' on page \thepage \space undefined}}%
    {\setbox\z@\hbox{\global\@tempcntc0\csname b@\@citeb\endcsname\relax}%
     \ifnum\@tempcntc=\z@ \@citeo\@tempcntb\m@ne
       \@citea\def\@citea{,}\hbox{\csname b@\@citeb\endcsname}%
     \else \advance\@tempcntb\@ne \ifnum\@tempcntb=\@tempcntc
      \else\advance\@tempcntb\m@ne\@citeo
      \@tempcnta\@tempcntc\@tempcntb\@tempcntc\fi\fi}}\@citeo}{#1}}
\def\@citeo{\ifnum\@tempcnta>\@tempcntb\else\@citea\def\@citea{,}%
  \ifnum\@tempcnta=\@tempcntb\the\@tempcnta\else
   {\advance\@tempcnta\@ne\ifnum\@tempcnta=\@tempcntb \else
   \def\@citea{--}\fi
   \advance\@tempcnta\m@ne\the\@tempcnta\@citea\the\@tempcntb}\fi\fi}
   \catcode`@=12

\begin{document}

\begin{flushright}
THES-TP/97-09\\
MPI/PhT/97-75\\ 
hep-ph/9711420\\
November 1997
\end{flushright}

\begin{center}
{\Large {\bf Higgs-Boson Low-Energy Theorem and }}\\[0.4cm] 
{\Large {\bf Compatible Gauge-Fixing Conditions}}\\[2.5cm]
{\large Apostolos Pilaftsis}\footnote[1]{E-mail address:
  pilaftsi@mppmu.mpg.de}\\[0.4cm] 
{\em Department of Theoretical Physics, University of Thessaloniki,}\\
{\em GR 54006 Thessaloniki, Greece}\\[0.3cm]
{\em and}\\[0.3cm]
{\em Max-Planck-Institut f\"ur Physik, F\"ohringer Ring 6, 
                                          80805 Munich, Germany}
\end{center}
\vskip1.7cm \centerline{\bf   ABSTRACT}    
Conventional gauge-fixing schemes such as R$_\xi$ gauges may lead to a
violation of the Higgs-boson low-energy theorem beyond the tree level.
To  elucidate  this fact,  we  study a simple  model  whose U(1) gauge
symmetry  is  spontaneously  broken,   and show   how the  Higgs-boson
low-energy theorem can consistently be extended to the gauge and Higgs
sectors of the model.  In this formulation, any gauge-fixing condition
must comply with the requirement that it  should be independent of the
vacuum expectation value of the Higgs field in  the symmetric limit of
the  theory.   We  give a    diagrammatic proof   of the   Higgs-boson
low-energy theorem to all orders   in perturbation theory, within  the
context of  a judiciously modified  R$_\xi$ gauge  compatible with the
above  constraint.  The dependence of  the kinematic parameters on the
Higgs tadpole is  found to  be very  important for  the proof of   the
theorem.

\newpage

The Higgs mechanism must be considered as the most natural solution to
the  problem of  generating the observed  masses for  the $W$  and $Z$
bosons  as well as for the  fermions,  {\em e.g.}, the electron, muon,
top  quark, {\em etc.}  Most  interestingly, such a mechanism does not
spoil other  desirable   field-theoretic properties  of  the quantized
action such  as unitarity and  renormalizability.   The latter is very
crucial in order that the  theory retains its predictive power  beyond
the Born  approximation.     The Higgs   mechanism  is  based  on  the
spontaneous symmetry breaking  (SSB) of a continuous (gauge) symmetry,
and reflects the fact that the true vacuum of the (Higgs) potential is
not rotationally invariant under  the continuous group.  Moreover, the
SSB of a global or  local group such  as the Standard Model (SM) gauge
group,  SU$(2)_L \otimes  $U(1)$_Y$,  gives rise  to  a massive scalar
particle, known as  the   Higgs boson ($H$), which,  however,  remains
elusive experimentally up to now.

Notwithstanding our poor experimental information, several theoretical
issues  have been  studied thus far  which are  closely related to the
nature  of the Higgs   scalar $H$.  In particular, Higgs  interactions
respect low-energy  theorems    \cite{EGN,SVZ,VVSS,VZS,DH,GHKD,KS,CKS}
analogous  to soft-pion theorems  in  hadron physics.   These theorems
relate Green functions   of   two transitions which  differ  from  one
another  by  the insertion of   a   Higgs boson with   zero momentum.  
Specifically, the Higgs-boson  low-energy  theorem (HLET) in its  most
basic form states that
\begin{equation}
  \label{HLET}
\lim_{p_H\to 0}\, \Gamma^{HAB}(p_H,p_A,p_B)\ =\
\frac{\partial}{\partial v}\, \Gamma^{AB}(p_A,-p_A)\, , 
\end{equation}
where all momenta   of  the generic  particles  $H$, $A$  and $B$  are
defined as  incoming ($p_H + p_A +  p_B  = 0$), and  $\Gamma^{AB}$ and
$\Gamma^{HAB}$ are the  two- and three-point correlation functions for
the transitions  $A\to  B$ and  $HA\to  B$, respectively.\footnote[1]{
  Notice that  the HLET  in Eq.\   (\ref{HLET}) is written  in a  more
  simplified but  equivalent  form to that given  in  the literature.  
  See, {\em   e.g.}, Ref.\ \cite{DH}.}    Beyond the  tree level,  all
kinematic parameters, including the vacuum expectation value (VEV) $v$
of the  Higgs  field,    must be   considered  as bare    quantities.  
Furthermore, the explicit dependence of the  bare masses and couplings
on  the    Higgs  tadpole  should  be  taken     into  account  on the
right-hand-side (RHS) of Eq.\ (\ref{HLET}). The relevant counter-terms
(CT's) required for  renormalization   may be derived   from low-order
correlation  functions, by  making use again   of the  HLET.  In  this
formulation, the differentiation   with respect to   $v$ acts on  {\em
  both} masses and mass-dependent couplings.  We will illuminate these
points  while  discussing a simple   ungauged model whose  U(1) global
symmetry is spontaneously broken.

One should  now observe that Eq.\  (\ref{HLET}) relates  amplitudes of
{\em physical} on-shell transitions {\em only} when the Higgs boson is
assumed to be massless, {\em i.e.}, $p^2_H =  M^2_H = 0$.  As we shall
see however, the requirement that the Higgs boson should be treated as
a massless particle is not a compelling  condition for the validity of
the   HLET.  In fact,  off-shell    transition amplitudes may be   the
sub-amplitudes of  high-loop   graphs,  so imposing  the    above mass
condition  may jeopardize the HLET at  high orders.  In this paper, we
wish  to extend the  formulation  of the HLET  to  the gauge and Higgs
sectors of a  SSB model.  Such a  consideration turns out to be highly
non-trivial, since conventional  gauge-fixing schemes  such as R$_\xi$
gauges can   invalidate  the equality  (\ref{HLET})   beyond the  Born
approximation.   This is not  very    surprising, as the   three-point
correlation function  $\Gamma^{HAB}(0,p_A,p_B)$   cannot represent  an
on-shell transition  for  massive   Higgs   bosons,  and is  hence   a
gauge-dependent quantity.  Within the  framework  of an Abelian  Higgs
model based on the SSB of the U(1)  gauge symmetry, we will explicitly
demonstrate the   above problem by means of   an example.  Finally, we
shall  show   diagrammatically how the  validity  of  the HLET  can be
maintained  to all orders  of  perturbation theory if the gauge-fixing
conditions are taken  to be independent of  the VEV of the Higgs field
in the unbroken limit of the Abelian Higgs model.

We start our discussion by considering an  Abelian ungauged model with
one  complex scalar  (Higgs)  field $\Phi$  and  one  fermion $f$. The
Lagrangian of the model is given by
\begin{equation}
  \label{Lscalar}
{\cal L}\ =\ (\partial_\mu \Phi^*)(\partial^\mu \Phi)\, 
+\, \bar{f}\, i\hspace{-1.2mm}\not\! \partial f\, 
-\, \kappa \bar{f}_L\Phi f_R\, 
-\, \kappa^* \bar{f}_R\Phi^* f_L\, +\, {\cal L}_V\, ,
\end{equation}
where ${\cal L}_V$ is the Higgs potential
\begin{equation}
  \label{Lpot}
{\cal L}_V\ =\  \mu^2 \Phi^*\Phi\, +\, \lambda (\Phi^*\Phi)^2\, . 
\end{equation}
The Lagrangian  is invariant under the global U(1) transformations:
\begin{equation}
  \label{trafoS}
\Phi\, \to \, e^{i\alpha} \Phi\, ,\qquad f_L\, \to \, e^{i\alpha}
f_L\, ,\qquad   f_R\, \to  f_R\, .
\end{equation}
The parameters $\mu$ and   $\lambda$ in Eq.\ (\ref{Lscalar}) are  real
numbers,  while $\kappa$ can always   be chosen real by performing  an
appropriate U(1) redefinition of the   left-handed fermion $f_L$.   If
$\lambda$ is  negative,  the global  U(1)  symmetry gets spontaneously
broken and the complex field $\Phi$  acquires a non-vanishing VEV $v$. 
The  Higgs field  must then be  expanded around  its VEV,  {\em i.e.},
$\Phi =  (v + H  + iG)/\sqrt{2}$.  The field $H$  is a massive CP-even
scalar  particle, the Higgs boson,  whereas $G$ is the massless CP-odd
Goldstone boson  associated with the SSB of  the global U(1) symmetry. 
The VEV  of $\Phi$ may be  determined by the minimization condition of
the Higgs potential
\begin{equation}
  \label{tadpole}
\frac{\partial\, {\cal L}_V}{\partial H}\, \Big|_{\langle \Phi\rangle}\
\equiv\ T\ =\ v(\mu^2\, +\, \lambda v^2)\, ,  
\end{equation}
with $T =  0$  and $v\not=  0$  at the tree  level.   Beyond the  Born
approximation, the tadpole  condition $T$ must be  adjusted in  such a
way such that the tadpole contribution to the Higgs boson which shifts
the  true vacuum must cancel.   It is therefore  important to keep the
explicit  dependence  of the  kinematic  parameters  on $T$.  Here, we
should also stress  that $T$ must  be  treated as bare quantity  whose
renormalized value is zero.   Finally, we remark that $\partial  {\cal
  L}_V  / \partial  G|_{\langle  \Phi\rangle} = 0$  which  is merely a
manifestation of the fact that $G$ represents the true Goldstone boson
of the theory.

The  bare parameters $\mu^0$,  $\lambda^0$,  $v^0$ and $T$, denoted by
the superscript `0', are not all independent of each other.  It proves
more convenient to express  the Lagrangian (\ref{Lscalar}) in terms of
$\mu^0$, $v^0$ and $T$, {\em i.e.},
\begin{equation}
  \label{lambda}
\lambda^0\ =\ -\, \frac{1}{(v^0)^2}\, \Big[ (\mu^0)^2\, -\, 
                                              \frac{T}{v^0}\, \Big]\, .
\end{equation}
After the SSB    of the U(1)  symmetry, the   bare Lagrangian may   be
expressed as
\begin{equation}
  \label{SSBscal}
{\cal L}^0\ =\ {\cal L}^0_{\rm kin}\, +\, {\cal L}^0_Y\, +\, {\cal L}^0_V\, ,
\end{equation}
where 
\begin{eqnarray}
  \label{L0kin}
{\cal L}^0_{\rm kin} & = &  \frac{1}{2}\, (\partial_\mu H^0)
(\partial^\mu H^0)\, +\,\frac{1}{2}\, (\partial_\mu G^0)(\partial^\mu G^0)\, 
+\, \bar{f}^0\, i\hspace{-1.2mm}\not\! \partial f^0\, ,\\
  \label{L0Y}
{\cal L}^0_Y & = & -\, \frac{\kappa^0v^0}{\sqrt{2}}\, 
\bar{f}^0\, \Big( 1\, +\, \frac{H^0}{v^0}\,
+\, i\gamma_5 \frac{G^0}{v^0}\, \Big)\, f^0\, ,\\
  \label{L0V}
{\cal L}^0_V & = & TH^0\, +\, \frac{T}{2v^0}\, (G^0)^2\, +\,
\frac{1}{2}\, \Big[ -2(\mu^0)^2\, +\, \frac{3T}{v^0}\, \Big]\,
(H^0)^2 \nonumber\\ 
&&+\, \frac{1}{2v^0}\, \Big[ -2(\mu^0)^2\, +\, \frac{2T}{v^0}\, \Big]\,
H^0 [(H^0)^2+ (G^0)^2]\nonumber\\
&&+\, \frac{1}{8(v^0)^2}\, \Big[ -2(\mu^0)^2\, +\, \frac{2T}{v^0}\, \Big]\,
[(H^0)^2+ (G^0)^2]^2\, .
\end{eqnarray}
Evidently, the Higgs  mechanism gives  rise to  a massive  fermion $f$
with $m^0_f = (\kappa^0v^0)/\sqrt{2}$ and  a massive Higgs scalar with
$(M^0_H)^2 = 2(\mu^0)^2 -  3T/v^0$.  The Goldstone boson  acquires the
squared mass  $(M^0_G)^2   =  -T/v^0$  proportional  to   the  tadpole
parameter $T$ and so vanishes after renormalization  to all orders, as
it should on account of the Goldstone theorem.

At this  point, it is  important  to notice that  the bare  parameters
$\mu^0$, $\lambda^0$   and $\kappa^0$ which   occur  in the  symmetric
formulation of  the  U(1) model  are   completely independent of   the
tadpole condition $T$. Instead, the  bare VEV $v^0$ depends implicitly
on $T$ through the relation (\ref{tadpole}). At one loop, for example,
the VEV may be given by
\begin{equation}
  \label{VEV}
v^0\ =\ v\, +\, \delta v\ =\ 
                  \Big(-\, \frac{(\mu^0)^2}{\lambda^0}\,\Big)^{1/2}\, -\, 
\frac{T}{M^2_H}\ ,
\end{equation}
with $M^2_H$ and $v$  denoting renormalized quantities. Thus,  the VEV
CT  $\delta v$ induces a contribution  of the tadpole parameter $T$ to
the Higgs and fermion self-energies through the second term on the RHS
of Eq.\ (\ref{VEV}).

For  notational  simplicity,   we drop  the  superscript  `0'  in  the
definition   of  bare  parameters in   the    following, unless it  is
explicitly stated otherwise.  The Feynman rules  of our ungauged  U(1)
model are displayed in Fig.\ 1.  Even  though $T$ as  well as $M_G$ is
zero  at   the  tree level, this   needs   not to  be   true for their
derivatives with respect to  $v$.  For  the  same exactly reasons,  we
must keep the full dependence of the bare kinematic parameters on $T$,
when calculating the RHS  of the HLET in  Eq.\ (\ref{HLET}).  In fact,
we have the following elementary identities:
\begin{eqnarray}
  \label{elids1}
\frac{\partial}{\partial v}\, m_f& =& \frac{m_f}{v}\ \equiv\ 
-\,\Gamma_0^{Hff}\, ,\qquad
\frac{\partial}{\partial v}\, M^2_H\ =\ \frac{3}{v}\, (M^2_H-M^2_G)\
\equiv\ -\Gamma_0^{HHH}\, ,\nonumber\\   
\frac{\partial}{\partial v}\, T &=& \mu^2 + 3\lambda v^2\ \equiv\
-M^2_H\, ,\qquad
\frac{\partial}{\partial v}\, M^2_G\ =\ \frac{1}{v}\, (M^2_H-M^2_G)\
\equiv\ -\Gamma_0^{HGG}\, ,\\
  \label{elids2} 
\frac{\partial}{\partial v}\, \Gamma_0^{HHH} &=& \Gamma_0^{HHHH}\,,
\qquad \frac{\partial}{\partial v}\, \Gamma_0^{HGG}\ =\
\Gamma_0^{HHGG}\,,\nonumber\\
\frac{\partial}{\partial v}\, \Gamma_0^{Hff} & =& 
\frac{\partial}{\partial v}\, \Gamma_0^{Gff}\ =\
\frac{\partial}{\partial v}\, \Gamma_0^{HHHH}\ =\ 
\frac{\partial}{\partial v}\,\Gamma_0^{HHGG}\ =\
\frac{\partial}{\partial v}\,\Gamma_0^{GGGG}\ =\ 0\, , 
\end{eqnarray}
where    the    tree-level   interaction vertices    $\Gamma_0^{Hff}$,
$\Gamma_0^{HHH}$,  $\Gamma_0^{HGG}$, {\em etc.}, may  be read off from
Fig.\ 1.    Note that all  the  above  tree-level identities  in Eqs.\ 
(\ref{elids1}) and (\ref{elids2})  are in complete accordance with the
HLET.  The very  same identities   play  an important role to   extend
diagrammatically the proof of the HLET  to higher orders. To this end,
one must also observe how an  insertion of a zero-momentum Higgs boson
occurs when  one differentiates the bare  $f$, $H$ and $G$ propagators
with respect to $v$, {\em i.e.},
\begin{eqnarray}
  \label{HLETprop}
\frac{\partial}{\partial v}\, iS_f(\not\! p) &=& iS_f(\not\! p)\, 
i\Gamma_0^{Hff}\, iS_f(\not\! p)\, ,\nonumber\\
\frac{\partial}{\partial v}\, i\Delta_H (p) &=& i\Delta_H (p)\, 
i\Gamma_0^{HHH}\, i\Delta_H (p)\, ,\nonumber\\
\frac{\partial}{\partial v}\, iD_G(p) &=& iD_G (p)\, 
i\Gamma_0^{HGG}\, iD_G (p)\, ,  
\end{eqnarray}
where   $S_f  = (\not\!    p -  m_f)^{-1}$,  $\Delta_H (p)   = (p^2  -
M^2_H)^{-1}$ and $D_G (p) = (p^2 - M^2_G)^{-1}$.  Eqs.\ (\ref{elids2})
and (\ref{HLETprop}) are  sufficient  to warranty the validity  of the
HLET  to all orders in  perturbation theory. Especially, with the help
of these equations, we can also understand how the $v$-derivative acts
on a high-order self-energy graph, $\Pi^{AA}$, with $A=f, H, G$.  Note
that  the  CP-violating  $HG$ mixing   is   completely absent in   the
CP-invariant U(1) model under discussion.

For illustration, we  give an one-loop  example which will help us  to
demonstrate explicitly how one can inductively show the HLET to higher
orders. Let  us consider the   transition amplitude $G(p)\to G(p)$  to
one-loop
\begin{equation}
  \label{GG}
\Gamma^{GG} (p,-p)\ =\ p^2\, -\, M^2_G\,  +\, \Pi^{GG}(p)\, .
\end{equation}
The unrenormalized $G$   self-energy $\Pi^{GG}(p)$  may be  decomposed
into two terms: one depending on the fermion $f$, $\Pi^{GG}_{(f)}(p)$,
and one  containing purely bosonic contributions, $\Pi^{GG}_{(b)}(p)$. 
Their explicit analytic form may be obtained by
\begin{eqnarray}
  \label{PiGGf}
\Pi^{GG}_{(f)}(p) &=& (-1)\times \int\!\! \frac{d^nk}{(2\pi)^n i}\,
{\rm Tr}\, \Big[\, i\Gamma^{Gff}_0\, iS_f(\not\! k)\, i\Gamma^{Gff}_0\,
iS_f(\not\! k - \not\! p)\, \Big]\, ,\\
\Pi^{GG}_{(b)}(p) &=& \int\!\! \frac{d^nk}{(2\pi)^n i}\, \Big[\, 
(i\Gamma^{HGG}_0)^2\, iD_G(k)\, i\Delta_H (k-p)\ +\
\frac{1}{2}\, i\Gamma^{HHGG}_0\, iD_G(k)\nonumber\\
&& +\  \frac{1}{2}\, i\Gamma^{HHHH}_0\, i\Delta_H (k)\, \Big]\, ,
\end{eqnarray}
where the loop  integrals must be evaluated  in $n = 4 - 2\varepsilon$
dimensions.  One should  bear  in mind that   $\Pi^{GG}(p)$ represents
one-particle irreducible   (1PI)   one-loop amplitude,   while tadpole
contributions, denoted as $\Gamma^H (0)$, enter via $M_G$. In fact, we
have $M^2_G = - T/v$, and $T$ may be derived by the condition
\begin{eqnarray}
T\, +\, \Gamma^H &=& 0\, ,\nonumber\\
  \label{GamHt}
T_{(f)} &=& -(-1)\times \int\!\! \frac{d^nk}{(2\pi)^n i}\,
{\rm Tr}\, \Big[\, i\Gamma^{Hff}_0\, iS_f(\not\! k)\,  \Big]\, ,\\
  \label{GamHb}
T_{(b)} &=& -\, \frac{1}{2} \int\!\! \frac{d^nk}{(2\pi)^n i}\, \Big[\, 
i\Gamma^{HGG}_0\, iD_G(k)\ +\ i\Gamma^{HHH}_0\, i\Delta_H (k)\, \Big]\, ,
\end{eqnarray}
where $T$  has also  been  written as a sum   of fermionic and bosonic
contributions. It is  a matter  of simple  algebra  to prove that  the
Goldstone boson $G$  remains massless to  one-loop after including the
tadpole graphs. Indeed, we find that 
\begin{equation}
  \label{Gmass0}
M^2_G - \Pi^{GG} (0)\ =\  - T/v\, -\, \Pi^{GG} (0)\ =\ 0\, ,
\end{equation}
in  agreement with  the Goldstone theorem.   If  we  now differentiate
$\Gamma^{GG}(p,-p)$ with respect to the VEV $v$ and use the elementary
identities  in  Eqs.\  (\ref{elids1})--(\ref{HLETprop}),  it  is  then
straightforward to obtain
\begin{equation}
  \label{GGLET}
\frac{\partial}{\partial v}\, \Gamma^{GG}(p,-p)\ =\ \Gamma_0^{HGG}\, +\,
\Gamma_1^{HGG}(0,p,-p)\, ,
\end{equation}
where $\Gamma_1^{HGG} = \Gamma_{1(f)}^{HGG} + \Gamma_{1(b)}^{HGG}$ with
\begin{eqnarray}
  \label{HGGf}
\Gamma_{1(f)}^{HGG}(0,p,-p) &=& (-1)\times  \int\!\! \frac{d^nk}{(2\pi)^n
  i}\, {\rm Tr}\, \Big[\, i\Gamma^{Gff}_0\, iS_f(\not\! k)\, i\Gamma^{Hff}_0\,
 iS_f(\not\! k)\, i\Gamma^{Gff}_0\, iS_f(\not\! k - \not\!  p)\nonumber\\ 
&&+\ i\Gamma^{Gff}_0\, iS_f(\not\! k)\, i\Gamma^{Gff}_0\,
iS_f(\not\! k - \not\! p)\, i\Gamma^{Hff}_0\,iS_f(\not\! k - \not\!  p)\, 
\Big]\, ,\\
\Gamma_{1(b)}^{HGG}(0,p,-p) &=&  \int\!\! \frac{d^nk}{(2\pi)^n i}\ \Big[\, 
2(i\Gamma^{HGG}_0)\, iD_G(k)\, i\Delta_H (k-p)\nonumber\\
&& +\
(i\Gamma^{HGG}_0)^2\, iD_G(k)\,i\Gamma^{HGG}_0\, iD_G(k)\, i\Delta_H (k-p)
\nonumber\\
&& +\ (i\Gamma^{HGG}_0)^2\, iD_G(k)\, i\Delta_H (k-p)\, i\Gamma^{HHH}_0\, 
i\Delta_H (k-p) \nonumber\\
&&+\
\frac{1}{2}\, i\Gamma^{HHGG}_0\, iD_G(k)\,i\Gamma^{HGG}_0\, iD_G(k)\nonumber\\
&& +\  \frac{1}{2}\, i\Gamma^{HHHH}_0\, i\Delta_H (k)\,
i\Gamma^{HHH}_0\, i\Delta_H (k)\,  \Big]\, .
\end{eqnarray}
Evidently, $\Gamma_1^{HGG}(0,p,-p)$ is the unrenormalized 1PI one-loop
$HGG$ coupling  evaluated with vanishing  Higgs-boson  momentum.  As a
consequence,  Eq.\ (\ref{GGLET}) proves the validity  of the HLET when
applied to  the $G$-boson vacuum  polarization. Apart from the $H$ and
$G$ wave-function renormalization constants which should be taken into
account in the calculation  of physical matrix  elements and  as usual
must  be  supplied   by  hand,  all   other  CT's  relevant   for  the
renormalization of the one-loop $HGG$ vertex are entirely contained in
$\Gamma_0^{HGG}$, namely,  in the lower-order (tree-level) correlation
function. 

The HLET can be applied equally well to three-,  four-, and all higher
$n$- point correlation  functions  at one  loop, and  checked  for its
validity  in an  exactly similar  manner.  The  proof of the  HLET can
inductively be carried over  to all orders. Specifically, the one-loop
amplitudes will be the  sub-amplitudes of two-loop graphs  and satisfy
relations very analogous  to Eqs.\ (\ref{elids1}) and  (\ref{elids2}),
where the  tree-level   correlation functions  are  replaced  by their
one-loop counterparts.    With   the  help  of  the  newly    obtained
identities,  the two-loop amplitudes can  be shown to respect the HLET
by  performing a diagrammatic analysis   very similar to the  one-loop
case.  Then, the two-loop amplitudes  obeying analogous identities  to
Eqs.\ (\ref{elids1}) and (\ref{elids2}) will  be the sub-amplitudes of
three-loop graphs and so on.

Our next consideration is to promote the above global U(1) symmetry of
the SSB model to a local symmetry, and discuss the consequences of the
gauge-fixing and the so-induced   ghost  terms on Eq.\  (\ref{HLET}).  
Before the SSB, the gauge-invariant part of the Lagrangian reads:
\begin{equation}
  \label{Lgauge}
{\cal L}_{\rm inv}\ =\ -\frac{1}{4}\, F_{\mu\nu} F^{\mu\nu}\, +\,
(D_\mu \Phi)^* (D^\mu \Phi)\, 
+\, \bar{f}_L\, i\hspace{-1mm}\not\!\! D f_L\, +\, 
\bar{f}_R\, i\hspace{-1.2mm}\not\! \partial f_R\,  
+\, {\cal L}_Y\, +\, {\cal L}_V\, ,
\end{equation}
where $F_{\mu\nu} = \partial_\mu A_\nu - \partial_\nu A_\mu$, $D_\mu =
\partial_\mu - i g A_\mu \widehat{Y}/2$, and $g$ and $\widehat{Y}$ are
the coupling constant and  the hypercharge generating operator  of the
local U(1)$_Y$  group, respectively.  To  avoid the  Adler-Bell-Jackiw
(ABJ) anomaly  \cite{ABJ}, one has  to introduce two fermions at least
with opposite    hypercharges, {\em i.e.},   $f =  (f_1, f_2)$.  To be
precise, the hypercharge  quantum numbers of  the different fields are
$Y_\Phi = 1$,  $Y_{1L} =1$ for $f_{1L}$, $Y_{2L}  = -1$ for  $f_{2L}$,
and $Y_{1R} =  Y_{2R} = 0$ for  $f_{1R}$ and $f_{2R}$.   Consequently,
under the local U(1)$_Y$ group, the fields transform as follows:
\begin{equation}
\label{trafoG}
\Phi \to \exp \Big[i\frac{Y_\Phi}{2}\, \theta(x)\Big]\, \Phi\, ,\quad
f \to \exp \Big[i\frac{Y_f}{2}\, \theta(x)\Big]\, f\, ,\quad
A_\mu \to A_\mu\, +\, \frac{1}{g}\, \partial_\mu\theta(x)\, .     
\end{equation}
Furthermore, the Lagrangian part of  the Higgs potential ${\cal  L}_V$
is identical to that given    in Eq.\ (\ref{Lpot}), while the   Yukawa
sector takes on the form
\begin{equation}
  \label{YukG}
{\cal L}_Y\ =\ -\, \bar{f}_L\, \left(\begin{array}{cc}
\kappa_1 \Phi & 0 \\
0 & \kappa_2 \Phi^* \end{array} \right)\, f_R \quad +\quad {\rm H.c.}, 
\end{equation}
where the Yukawa couplings  $\kappa_1$ and $\kappa_2$  can be taken to
be real  numbers, {\em i.e.}, we assume  absence of mixing between the
two fermions.

To remove the unphysical degrees of  freedom from the gauge field, one
has to break   the continuous U(1)$_Y$  gauge  symmetry by introducing
gauge-fixing (GF) and Fadeev-Popov (FP) ghost terms, denoted as ${\cal
  L}_{\rm GF}$ and  ${\cal   L}_{\rm  FP}$, respectively. Then,    the
quantized Lagrangian ${\cal  L} = {\cal  L}_{\rm inv} + {\cal  L}_{\rm
  GF} + {\cal L}_{\rm FP}$ is invariant under Becchi-Rouet-Stora (BRS)
transformations \cite{BRS}. For reasons that  will become more obvious
later on, we adopt the R$_\xi$-type GF condition
\begin{equation}
  \label{GFcond}
F(x)\ =\ \partial_\mu A^\mu\,  +\, \frac{g}{2}\, \xi\, \Im m(\Phi^2)\, ,
\end{equation}
whence the GF Lagrangian reads:
\begin{equation}
  \label{LGF}
{\cal  L}_{\rm  GF}\ =\ -\, \frac{1}{2\xi}\, \Big[\, \partial_\mu A^\mu\,
 +\, \frac{g}{2}\, \xi\, \Im m(\Phi^2)\, \Big]^2\, , 
\end{equation}
where   $\xi$ is   the gauge-fixing    parameter  (GFP). Imposing  BRS
invariance on the Lagrangian gives rise to FP ghost interactions which
are obtained by
\begin{equation}
  \label{LFP}
{\cal L}_{\rm FP}\ =\ -\, g\, \bar{c}\, \frac{\delta F}{\delta \theta}\,
c\ =\ - \bar{c}\, \Big[\, \partial_\mu\partial^\mu\, +\, \frac{g^2}{2}\,
\xi\, \Re e(\Phi^2)\, \Big]\, c\, ,
\end{equation}
where  $c$  and $\bar{c}$  are   complex FP  scalars.   As usual,  the
variation of the GF condition $F(x)$ in Eq.\ (\ref{LFP}) is taken over
the  local  gauge  transformations  of $\Phi$   and $A_\mu$   in  Eq.\ 
(\ref{trafoG}).

After the  SSB of the Higgs potential  ${\cal L}_V$,  the field $\Phi$
must be  expanded around its VEV  $v$, in exactly  the same way we did
for   the ungauged   scalar model  discussed    above.  Note  that the
minimization condition $T$ in     Eq.\ (\ref{tadpole})  is    entirely
determined  from ${\cal  L}_V$,  and one should   not include explicit
U(1)$_Y$  breaking terms  from  the unphysical  sector ${\cal  L}_{\rm
  GF}$, which  is designed so as  to cancel the  unphysical degrees of
freedom of the gauge field $A_\mu$ in ${\cal  L}_{\rm inv}$. The Higgs
mechanism gives rise  to  a massive  gauge field $A_\mu$  with  $M_A =
(gv)/2$, two  massive fermions $f_1$  and $f_2$ with $m_1  = (\kappa_1
v)/\sqrt{2}$ and $m_2 =   (\kappa_2 v)/\sqrt{2}$, and the Higgs  boson
with a  mass equal to that found  in the ungauged   scalar model.  The
would-be  Goldstone boson $G$ is eaten  by  the longitudinal degree of
the gauge boson $A_\mu$, and is  therefore unphysical; its mass is GFP
dependent, and it  decouples    from  S-matrix elements.    The   free
propagators  of the gauge boson, the  would-be Goldstone boson and the
ghost field are respectively given by
\begin{eqnarray}
  \label{Aprop}
\Delta_A^{\mu\nu} (p) &=& U^{\mu\nu} (p)\, -\, \frac{p^\mu
  p^\nu}{M^2_A}\, D_c (p)\, ;\quad   U^{\mu\nu} (p)\ =\ \Big(
  -g^{\mu\nu}\, +\, \frac{p^\mu p^\nu}{M^2_A}\, \Big)\, (p^2\,
  -\, M^2_A)^{-1}\, ,\\
  \label{Gprop}
D_G (p) &=& (p^2\, -\, \xi M^2_A\, -\, M^2_G)^{-1}\, ,\\
  \label{cprop}
D_c (p) &=& (p^2\, -\, \xi M^2_A )^{-1}\, ,
\end{eqnarray}
with  $M^2_G = -  T/v$.   The  fermion and  Higgs propagators  are not
modified by considering  the gauged version of  the U(1) scalar model. 
The remaining Feynman rules are shown in  Fig.\ 2. In the R$_\xi$-type
gauge considered here, $GA_\mu$ mixing is absent at the tree level. In
contrast to the conventional R$_\xi$ gauge, generated by the condition
\begin{equation}
  \label{Rxi}
F_{{\rm R}_\xi} (x)\ =\   \partial_\mu A^\mu\,  +\, \xi\, M_A G\, ,
\end{equation}
new  Feynman rules occur in  the gauge  defined in Eq.\ (\ref{GFcond})
which are very crucial for maintaining the HLET  through all orders as
we shall see below.

{}From   Eqs.\ (\ref{Gprop}) and  (\ref{cprop}), it  is worth noticing
that the bare  Goldstone and ghost  propagators are different  in this
formulation; they differ by the term $M^2_G$.  The latter term is very
important since  the  variation of the $G$  and  $c$  propagators with
respect to $v$ is  also different and  in complete agreement with Eq.\ 
(\ref{HLET}). More explicitly, we find  for the inverse $A_\mu$,  $G$,
and $c$ propagators that
\begin{eqnarray}
  \label{Gelid1}
\frac{\partial}{\partial v}\, \Delta_{A\mu\nu}^{-1} (p) &=& 
g_{\mu\nu}\, g\, M_A\ \equiv\ \Gamma_{0\mu\nu}^{HAA}\, ,\\
  \label{Gelid2}
\frac{\partial}{\partial v}\, D_G^{-1} (p) & = & -\,
\frac{g}{2M_A}\, (M^2_H - M^2_G + 2\xi M^2_A )\ \equiv\ \Gamma_0^{HGG}\, ,\\
  \label{Gelid3}
\frac{\partial}{\partial v}\, D_c^{-1} (p) & = & -\,
g\,\xi M_A \ \equiv\ \Gamma_0^{H\bar{c}c}\, .
\end{eqnarray}
We can also derive a wealth of elementary identities very analogous to
Eqs.\ (\ref{elids1}) and (\ref{elids2})  which obey Eq.\ (\ref{HLET}). 
It   is now interesting  to  see how the HLET   ceases to  hold in the
conventional  R$_\xi$ gauge at the  one-loop  level.  The gauge-fixing
condition $F_{{\rm R}_\xi}$ leads  to a GFP-independent $HGG$ coupling
equal  to   $-g^2M^2_H /  (2M_A)$    which  violates explicitly   Eq.\ 
(\ref{Gelid2}).   Also,  Eq.\  (\ref{Gelid3}) does   not respect  Eq.\ 
(\ref{HLET}) in the    usual  R$_\xi$ gauge, since  the    $H\bar{c}c$
coupling turns out to  be  short by a factor   of two in that gauge.   
Nevertheless,    Eq.\  (\ref{HLET})   is   satisfied    from both  the
gauge-fixing schemes mentioned above for the  gauge propagator and all
couplings  involving physical particles  at the tree level. Beyond the
Born approximation, even   the latter correlation  functions will  not
respect   the  HLET, since  the   tree-level $HGG$  and/or $H\bar{c}c$
vertices will be the sub-amplitudes of one-loop graphs.

It is now obvious that the actual reason for the  above failure of the
HLET resides in  the choice of  the gauge-fixing condition $F$ and its
dependence  on the VEV  $v$ of the field $\Phi$.  In order that $F$ be
compatible with the  HLET,  it should {\em  not}  depend  on $v$  when
considering  the unbroken  limit of the  gauge theory,  {\em i.e.}, it
should satisfy the condition
\begin{equation}
  \label{Comp}
\frac{\partial}{\partial v}\, F(x)\ =\ 0\, ,
\end{equation}
where all Higgs fields  are expressed in  terms of the unbroken fields
$\Phi$ and $\Phi^*$ as well as  linear combinations of them.  Clearly,
the gauge-fixing condition  in Eq.\ (\ref{GFcond}) satisfies this very
last criterion, whereas the conventional R$_\xi$ gauge,
\begin{equation}
  \label{FRxi}
F_{{\rm R}_\xi} (x)\ =\   \partial_\mu A^\mu\,  +\,
                         \frac{gv}{\sqrt{2}}\, \xi\, \Im m (\Phi )\, ,
\end{equation}
does not, {\em viz.}, $\partial F_{{\rm R}_\xi}/\partial v = g\, \xi\,
\Im  m (\Phi  )/\sqrt{2} \not=  0$.  One may  now  be  tempted to draw
interesting comparisons between the  known  Ward identities (WI's)  in
quantum electrodynamics (QED)  and      the  HLET stated     in   Eq.\ 
(\ref{HLET}).     In  QED, WI's   may   be  derived  by   the  minimal
substitution,  $\partial_\mu\to  \partial_\mu   -i   eA_\mu$,  of  the
four-momentum operator $\partial_\mu$.   By analogy,  the HLET may  be
obtained by the invariance of  the quantized Lagrangian in the  broken
phase under the  {\em subsequent} ``minimal'' shifts $H\to  H - v$ and
$v \to  v + H$ and vice  versa.  Obviously, $F_{{\rm R}_\xi}$ does not
possess the latter translational symmetry.

It is now rather instructive to analyze an one-loop example within the
U(1)$_Y$ gauge  model. For  this  purpose, we consider  the transition
amplitude $A_\mu (p) \to A_\nu (p)$, given by
\begin{equation}
  \label{AA}
\Gamma^{AA}_{\mu\nu}(p,-p)\ =\ t_{\mu\nu} \Gamma^{AA}_T (p)\, +\,
\ell_{\mu\nu} \Gamma^{AA}_L (p)\, ,  
\end{equation}
with $t_{\mu\nu} = -g_{\mu\nu} + p_\mu p_\nu / p^2$ and $\ell_{\mu\nu}
= p_\mu p_\nu / p^2$, and
\begin{eqnarray}
  \label{AAT}
\Gamma^{AA}_T (p) &=& p^2\, -\, M^2_A \, +\, \Pi^{AA}_T (p)\, ,\\
  \label{AAL}
\Gamma^{AA}_L (p) &=& \frac{1}{\xi}\, p^2\, -\, M^2_A \, +\, \Pi^{AA}_L (p)\, .
\end{eqnarray}
In Eqs.\ (\ref{AAT}) and (\ref{AAL}), $\Pi^{AA}_T (p)$ and $\Pi^{AA}_L
(p)$ are the transverse and longitudinal components of the $A$ vacuum
polarization, {\em i.e.},
\begin{equation}
  \label{PiAA}
\Pi^{AA}_{\mu\nu} (p) \ =\ t_{\mu\nu} \Pi^{AA}_T (p)\, +\, 
\ell_{\mu\nu} \Pi^{AA}_L (p)\, .   
\end{equation}
In the calculation of  $\Pi^{AA}_{\mu\nu} (p)$, we will omit fermionic
loops since they can easily be shown to satisfy Eq.\ (\ref{HLET}) very
similar to  the ungauged scalar model.   The 1PI  one-loop gauge-boson
self-energy is then written
\begin{eqnarray}
  \label{AAmunu}
\Pi^{AA}_{\mu\nu} (p) \!\!&=&\hspace{-8pt} 
\int \!\! \frac{d^nk}{(2\pi)^n i}\, \Big[\,
i\Gamma^{HAA}_{0\mu\lambda}\, i\Delta_A^{\lambda\rho} (k)\, 
i\Gamma^{HAA}_{0\rho\nu}\, i\Delta_H (p+k)\, 
+\,  i\Gamma^{HAG}_{0\mu}\, iD_G (k)\, 
i\Gamma^{HAG}_{0\nu}\, i\Delta_H (p+k)\, \nonumber\\
&&+\, \frac{1}{2}\, i\Gamma^{HHAA}_{0\mu\nu}\, i\Delta_H (k)\,
+\, \frac{1}{2}\, i\Gamma^{GGAA}_{0\mu\nu}\, iD_G (k)\, \Big]\, .
\end{eqnarray}
Employing     elementary   identities      analogous   to        Eqs.\ 
(\ref{elids1})--(\ref{HLETprop})  and  (\ref{Gelid1})--(\ref{Gelid3}),
it is      not     difficult  to    calculate  the    derivative    of
$\Gamma^{AA}_{\mu\nu} (p)$ with respect to $v$. In this way, we obtain
\begin{equation}
  \label{partAA}
\frac{\partial}{\partial v}\, \Gamma^{AA}_{\mu\nu} (p, -p)\ =\
\Gamma^{HAA}_{0\mu\nu}\, +\, \Gamma^{HAA}_{1\mu\nu} (0, p, -p)\, -\,
\delta_{\mu\nu} (p)\, ,
\end{equation}
where
\begin{eqnarray}
  \label{HdelAA}
\Gamma^{HAA}_{1\mu\nu}(0, p, -p)\, -\, \delta_{\mu\nu}(p)
&=& \int \!\! \frac{d^nk}{(2\pi)^n i}\,\Big[\, 
i\Gamma^{HHAA}_{0\mu\lambda}\, i\Delta_A^{\lambda\rho} (k)\, 
i\Gamma^{HAA}_{0\rho\nu}\, i\Delta_H (p+k)\nonumber\\
&&\hspace{-4cm}
+\, i\Gamma^{HAA}_{0\mu\lambda}\, i\Delta_A^{\lambda\rho} (k)\, 
i\Gamma^{HHAA}_{0\rho\nu}\, i\Delta_H (p+k)\, \nonumber\\
&&\hspace{-4cm}
+\, i\Gamma^{HAA}_{0\mu\lambda}\, i\Delta_A^{\lambda\sigma} (k)\, 
\Gamma^{HAA}_{0\sigma\tau}\, i\Delta_A^{\tau\rho} (k)\, 
i\Gamma^{HAA}_{0\rho\nu}\, i\Delta_H (p+k)\, \nonumber\\ 
&&\hspace{-4cm}
+\, i\Gamma^{HAA}_{0\mu\lambda}\, i\Delta_A^{\lambda\rho} (k)\, 
i\Gamma^{HAA}_{0\rho\nu}\, i\Delta_H (p+k)\, i\Gamma^{HHH}_0\,
i\Delta_H (p+k)\, \nonumber\\
&&\hspace{-4cm}
+\, i\Gamma^{HAG}_{0\mu}\, iD_G (k)\, i\Gamma^{HGG}_0\, iD_G (k)\, 
i\Gamma^{HAG}_{0\nu}\, i\Delta_H (p+k)\, \nonumber\\
&&\hspace{-4cm}
+\,  i\Gamma^{HAG}_{0\mu}\, iD_G (k)\, i\Gamma^{HAG}_{0\nu}\, 
i\Delta_H (p+k)\, i\Gamma^{HHH}_0\, i\Delta_H (p+k)\, \nonumber\\
&&\hspace{-4cm}
+\, \frac{1}{2}\, \Big( i\Gamma^{HHAA}_{0\mu\nu}\, i\Delta_H (k)\,
i\Gamma^{HHH}_0\, i\Delta_H (k)\, +\, 
i\Gamma^{GGAA}_{0\mu\nu}\, iD_G (k)\,
i\Gamma^{HGG}_0\, iD_G (k) \Big)\, \Big]
\end{eqnarray}
and 
\begin{eqnarray}
  \label{delmunu}
\hspace{-4cm}\delta_{\mu\nu}(p) 
&=& \int \!\! \frac{d^nk}{(2\pi)^n i}\,
\Big[\, i\Gamma^{HAA}_{0\mu\lambda}\, i\Delta_A^{\lambda\rho} (k)\, 
i\Gamma^{HAG}_{0\rho}\, iD_G (k)\, i\Gamma^{HAG}_{0\nu}\,
i\Delta_H (p+k) \, \nonumber\\
&&
+\, i\Gamma^{HAG}_{0\mu}\, iD_G (k)\, i\Gamma^{HAG}_{0\rho}\, 
i\Delta_A^{\rho\lambda} (k)\, i\Gamma^{HAA}_{0\lambda\nu}\,
i\Delta_H (p+k)\, \Big]\, .
\end{eqnarray}
In Eq.\ (\ref{partAA}), $\Gamma^{HAA}_{1\mu\nu} (0, p, -p)$ stands for
the unrenormalized 1PI   one-loop $HAA$ coupling evaluated with   zero
Higgs-boson    momentum, {\em   i.e.},  $p_H  =    0$.  The expression
$\delta_{\mu\nu}(p)$ quantifies the would-be  deviation from the HLET. 
However,   $\delta_{\mu\nu}(p)$   vanishes identically  in  the  limit
$p_H\to 0$   since it  is proportional  to   $\Gamma^{HAG}_{0\rho} = g
p_{H\rho}$.  In the usual  R$_\xi$  gauge, the situation  is different
since the $HAG$ coupling also depends on the Goldstone-boson momentum.
As a result, one finds that in the latter gauge $\delta_{\mu\nu} \not=
0$ which  explicitly breaks the HLET.  In  addition, we note  that the
lower-order correlation (tree-level) function $\Gamma^{HAA}_{0\mu\nu}$
in Eq.\  (\ref{partAA})   provides   the   necessary  CT's  for    the
renormalization  of  the  one-loop  $HAA$   vertex.  As   before,  the
wave-function renormalizations of the  external particles, {\em e.g.},
$Z_H$ and $Z_A$, should also be taken into account.

The proof of the HLET in the U(1)$_Y$ gauge model  with SSB may now be
extended by induction to all orders.   By making use of a diagrammatic
analysis similar to  that outlined above  for the ungauged U(1) model,
one  can show    that the   one-loop  correlation  functions   satisfy
identities  analogous to  their   tree-level counterparts in  complete
accordance with  the  HLET.  Then, the one-loop  correlation functions
will constitute the sub-amplitudes  of two-loop graphs, giving rise to
identities which in turn obey the HLET and so on.

In summary, we have explicitly demonstrated that the tadpole condition
$T$ plays a significant  role in extending  the validity of  the HLET,
stated in Eq.\ (\ref{HLET}),  to the gauge and Higgs  sectors of a SSB
model  such   as the  Abelian  Higgs   model  studied here.    In this
formulation, it is   important to know  how  all  particle masses  and
mass-dependent  couplings vary  under an  infinitesimal  shift of  the
Higgs  VEV $v$ or   equivalently how they  depend  on  $T$.   In gauge
theories with    SSB, there  is  the  additional  problem   that naive
gauge-fixing conditions $F$ may lead to a  violation of the HLET, {\em
  i.e.}, gauge-dependent terms violate  Eq.\ (\ref{HLET}). As has been
shown, $F$  must be independent on  $v$  in the  unbroken limit of the
theory in order that  the equality Eq.\  (\ref{HLET}) holds  to higher
orders.   Finally, it is  interesting to comment on other gauge-fixing
schemes, such as the background field method  (BFM) \cite{BFM}, and on
diagrammatic approaches  based on the pinch  technique (PT) \cite{PT}. 
In the BFM, gauge transformations of the background fields respect the
local symmetries of the gauge-invariant part  of the classical action. 
Furthermore, it  is  known that BFM  fields,  including the  BFM Higgs
boson,  appear only at   the  external legs  of $n$-point  correlation
functions.  Since the BFM gauge-fixing   condition is compatible  with
the  requirement  in Eq.\ (\ref{Comp}),  the  validity of  the HLET is
expected.  The very  same property  shares the  $n$-point  correlation
functions   evaluated  by the  PT.\footnote[1]{Explicit   calculations
  testing the HLET  in the PT may be  found in  Ref.\ \cite{KS}.}  The
main difference as well as advantage of the PT over other gauge-fixing
schemes is the fact  that the constraint  (\ref{Comp}) on $F$'s  is no
longer needed as soon as the pinching contributions are properly taken
into consideration in the   calculation of the effective   PT one-loop
functions.  Detailed study of the latter may be given elsewhere.

The  HLET reflects the underlying  gauge symmetry of  the theory after
the  latter is spontaneously  broken  by  the Higgs  mechanism.  As  a
consequence of the theorem, the interactions of the particles with the
Higgs boson   are  closely related to  their    observed masses.  This
constitutes a fundamental  property distinguishing the Higgs nature of
a fundamental scalar from other  particles, which will extensively  be
probed at the CERN Large Hadron Collider.

\bigskip\bigskip
\noindent
{\bf Acknowledgements.}  I   wish to thank  Bernd Kniehl  for  helpful
discussions and remarks.

\newpage


\centerline{\Large{\bf Figure captions }}
\vspace{-0.2cm}
\newcounter{fig}
\begin{list}{\rm {\bf Fig. \arabic{fig}: }}{\usecounter{fig}
\labelwidth1.6cm \leftmargin2.5cm \labelsep0.4cm \itemsep0ex plus0.2ex }

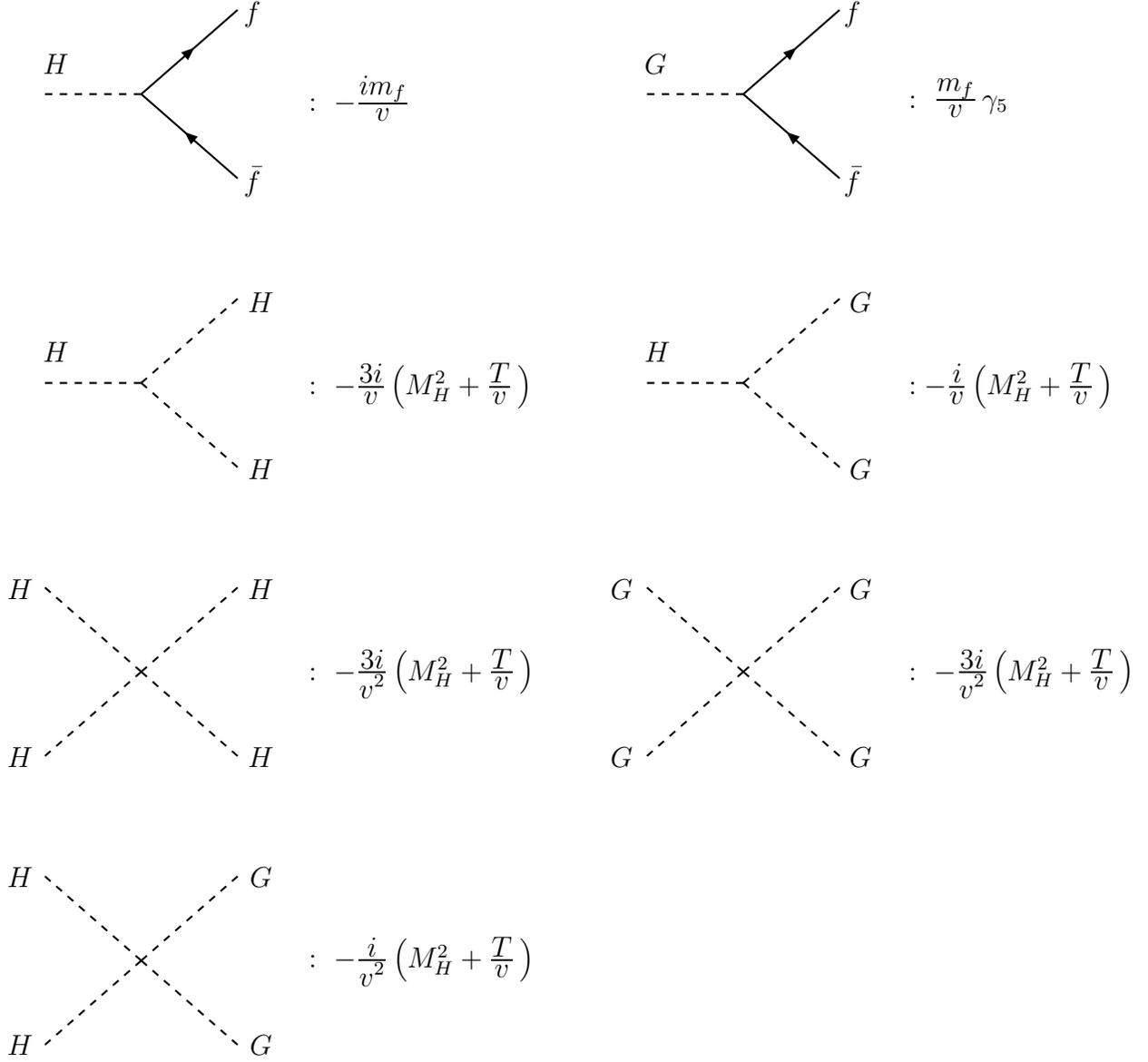
\item[{\bf Fig.\ 1:}] Feynman rules for the ungauged Abelian Higgs model. 

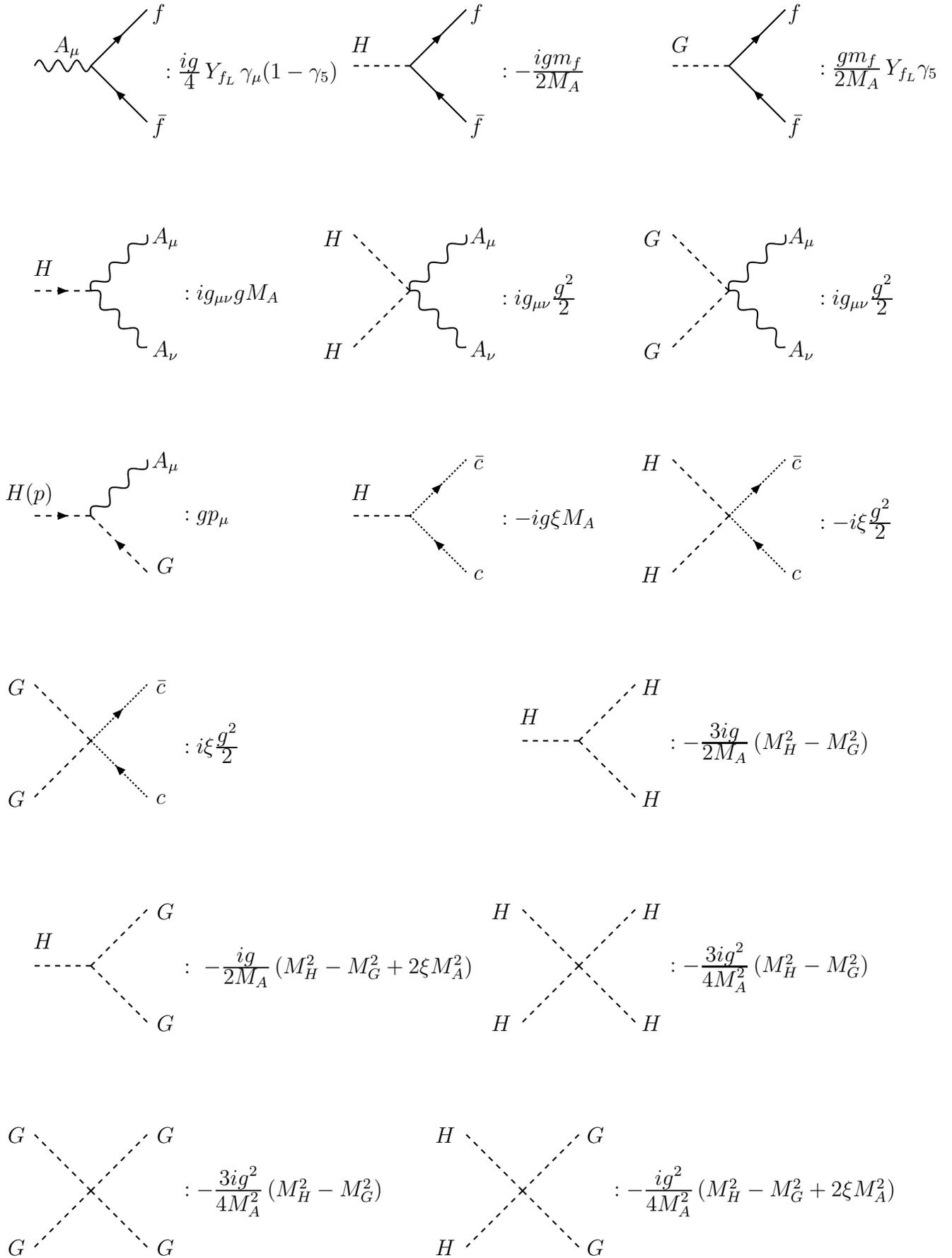
\item[{\bf Fig.\ 2:}] Feynman rules pertaining to the U(1)$_Y$ gauge
  model with SSB.

\end{list}

\newpage

\begin{center}
\begin{picture}(400,500)(0,0)
\SetWidth{0.8}

\DashLine(0,420)(40,420){3}
\ArrowLine(40,420)(80,455)
\ArrowLine(80,385)(40,420)
\Text(10,430)[rb]{$H$}
\Text(90,455)[r]{$f$}
\Text(90,385)[r]{$\bar{f}$}
\Text(110,420)[l]{$:\ -\frac{\displaystyle im_f}{\displaystyle v}$}


\DashLine(250,420)(290,420){3}
\ArrowLine(290,420)(330,455)
\ArrowLine(330,385)(290,420)
\Text(250,430)[lb]{$G$}
\Text(340,455)[r]{$f$}
\Text(340,385)[r]{$\bar{f}$}
\Text(360,420)[l]{$:\ \frac{\displaystyle m_f}{\displaystyle v}\,\gamma_5$}

\DashLine(0,300)(40,300){3}
\DashLine(40,300)(80,335){3}
\DashLine(80,265)(40,300){3}
\Text(10,310)[rb]{$H$}
\Text(90,335)[]{$H$}
\Text(90,265)[]{$H$}
\Text(110,300)[l]{$:\ -\frac{\displaystyle 3i}{\displaystyle v}\, 
\Big( M^2_H + \frac{\displaystyle T}{\displaystyle v}\, \Big)$}


\DashLine(250,300)(290,300){3}
\DashLine(290,300)(330,335){3}
\DashLine(330,265)(290,300){3}
\Text(250,310)[lb]{$H$}
\Text(340,335)[]{$G$}
\Text(340,265)[]{$G$}
\Text(360,300)[l]{$:-\frac{\displaystyle i}{\displaystyle v}\, 
\Big( M^2_H + \frac{\displaystyle T}{\displaystyle v}\, \Big)$}

\DashLine(0,215)(40,180){3}
\DashLine(0,145)(40,180){3}
\DashLine(40,180)(80,215){3}
\DashLine(40,180)(80,145){3}
\Text(-5,215)[r]{$H$}
\Text(-5,145)[r]{$H$}
\Text(90,215)[]{$H$}
\Text(90,145)[]{$H$}
\Text(110,180)[l]{$:\ -\frac{\displaystyle 3i}{\displaystyle v^2}\, 
\Big( M^2_H + \frac{\displaystyle T}{\displaystyle v}\, \Big)$}


\DashLine(250,215)(290,180){3}
\DashLine(250,145)(290,180){3}
\DashLine(290,180)(330,215){3}
\DashLine(290,180)(330,145){3}
\Text(245,215)[r]{$G$}
\Text(245,145)[r]{$G$}
\Text(340,215)[]{$G$}
\Text(340,145)[]{$G$}
\Text(360,180)[l]{$:\ -\frac{\displaystyle 3i}{\displaystyle v^2}\, 
\Big( M^2_H + \frac{\displaystyle T}{\displaystyle v}\, \Big)$}

\DashLine(0,95)(40,60){3}
\DashLine(0,25)(40,60){3}
\DashLine(40,60)(80,95){3}
\DashLine(40,60)(80,25){3}
\Text(-5,95)[r]{$H$}
\Text(-5,25)[r]{$H$}
\Text(90,95)[]{$G$}
\Text(90,25)[]{$G$}
\Text(110,60)[l]{$:\ -\frac{\displaystyle i}{\displaystyle v^2}\, 
\Big( M^2_H + \frac{\displaystyle T}{\displaystyle v}\, \Big)$}

\Text(200,0)[t]{\bf Fig.\ 1}

\end{picture}
\end{center}

\newpage

\begin{center}
\begin{picture}(455,640)(0,0)
\SetWidth{0.8}

\Photon(0,660)(30,660){3}{3}
\ArrowLine(30,660)(60,690)
\ArrowLine(60,630)(30,660)
\Text(10,670)[lb]{$A_{\mu}$}
\Text(70,690)[r]{$f$}
\Text(70,630)[r]{$\bar{f}$}
\Text(70,660)[l]{$: \frac{\displaystyle ig}{\displaystyle 4}\,
Y_{f_L}\,\gamma_{\mu}( 1 - \gamma_5)$}

\DashLine(170,660)(200,660){3}
\ArrowLine(200,660)(230,690)
\ArrowLine(230,630)(200,660)
\Text(180,670)[rb]{$H$}
\Text(240,690)[r]{$f$}
\Text(240,630)[r]{$\bar{f}$}
\Text(250,660)[l]{$: -\frac{\displaystyle igm_f}{\displaystyle 2M_A}$}

\DashLine(340,660)(370,660){3}
\ArrowLine(370,660)(400,690)
\ArrowLine(400,630)(370,660)
\Text(350,670)[rb]{$G$}
\Text(410,690)[r]{$f$}
\Text(410,630)[r]{$\bar{f}$}
\Text(420,660)[l]{$: \frac{\displaystyle gm_f}{\displaystyle 2M_A}\, 
Y_{f_L}\gamma_5$}

\DashArrowLine(0,540)(30,540){3}
\Photon(30,540)(60,570){3}{3} 
\Photon(30,540)(60,510){3}{3}
\Text(10,550)[rb]{$H$}
\Text(70,570)[]{$A_{\mu}$}
\Text(70,510)[]{$A_{\nu}$}
\Text(80,540)[l]{$: ig_{\mu\nu}gM_A$}

\DashLine(170,570)(200,540){3}
\DashLine(170,510)(200,540){3}
\Photon(200,540)(230,570){3}{3}
\Photon(200,540)(230,510){3}{3}
\Text(165,570)[r]{$H$}
\Text(165,510)[r]{$H$}
\Text(240,570)[]{$A_{\mu}$}
\Text(240,510)[]{$A_{\nu}$}
\Text(250,540)[l]{$: ig_{\mu\nu} \frac{\displaystyle
    g^2}{\displaystyle 2}$}

\DashLine(340,570)(370,540){3}
\DashLine(340,510)(370,540){3}
\Photon(370,540)(400,570){3}{3}
\Photon(370,540)(400,510){3}{3}
\Text(335,570)[r]{$G$}
\Text(335,510)[r]{$G$}
\Text(410,570)[]{$A_{\mu}$}
\Text(410,510)[]{$A_{\nu}$}
\Text(420,540)[l]{$: ig_{\mu \nu} \frac{\displaystyle
    g^2}{\displaystyle 2}$}

\DashArrowLine(0,420)(30,420){3}
\Photon(30,420)(60,450){3}{3} 
\DashArrowLine(60,390)(30,420){3}
\Text(10,430)[rb]{$H(p)$}
\Text(70,450)[]{$A_{\mu}$}
\Text(70,395)[]{$G$}
\Text(80,420)[l]{$: g p_{\mu}$}

\DashLine(170,420)(200,420){3}
\DashArrowLine(200,420)(230,450){1}
\DashArrowLine(230,390)(200,420){1}
\Text(180,430)[rb]{$H$}
\Text(240,450)[r]{$\bar{c}$}
\Text(240,390)[r]{$c$}
\Text(250,420)[l]{$: -ig \xi M_A$}

\DashLine(340,450)(370,420){3}
\DashLine(340,390)(370,420){3}
\DashArrowLine(370,420)(400,450){1}
\DashArrowLine(400,390)(370,420){1}
\Text(335,450)[r]{$H$}
\Text(335,390)[r]{$H$}
\Text(410,450)[r]{$\bar{c}$}
\Text(410,390)[r]{$c$}
\Text(420,420)[l]{$: -i \xi \frac{\displaystyle g^2}{\displaystyle 2}$}

\DashLine(0,330)(30,300){3}
\DashLine(0,270)(30,300){3}
\DashArrowLine(30,300)(60,330){1} 
\DashArrowLine(60,270)(30,300){1}
\Text(-5,330)[r]{$G$}
\Text(-5,270)[r]{$G$}
\Text(70,330)[r]{$\bar{c}$}
\Text(70,270)[r]{$c$}
\Text(80,300)[l]{$: i \xi \frac{\displaystyle g^2}{\displaystyle 2}$}

\DashLine(260,300)(290,300){3}
\DashLine(290,300)(320,330){3}
\DashLine(320,270)(290,300){3}
\Text(270,310)[rb]{$H$}
\Text(330,330)[]{$H$}
\Text(330,270)[]{$H$}
\Text(340,300)[l]{$: -\frac{\displaystyle 3ig}{\displaystyle 2M_A}
\, ( M^2_H-M^2_G )$}

\DashLine(0,180)(30,180){3}
\DashLine(30,180)(60,210){3}
\DashLine(60,150)(30,180){3}
\Text(10,190)[rb]{$H$}
\Text(70,210)[]{$G$}
\Text(70,150)[]{$G$}
\Text(80,180)[l]{$:\ -\frac{\displaystyle ig}{\displaystyle 2M_A}\,
( M^2_H-M^2_G+2\xi M^2_A )$}


\DashLine(260,210)(290,180){3}
\DashLine(260,150)(290,180){3}
\DashLine(290,180)(320,210){3} 
\DashLine(320,150)(290,180){3}
\Text(255,210)[r]{$H$}
\Text(255,150)[r]{$H$}
\Text(330,210)[]{$H$}
\Text(330,150)[]{$H$}
\Text(340,180)[l]{$: -\frac{\displaystyle 3ig^2}{\displaystyle
    4M^2_A}\, ( M^2_H - M^2_G )$}

\DashLine(0,90)(30,60){3}
\DashLine(0,30)(30,60){3}
\DashLine(30,60)(60,90){3}
\DashLine(30,60)(60,30){3}
\Text(-5,90)[r]{$G$}
\Text(-5,30)[r]{$G$}
\Text(70,90)[]{$G$}
\Text(70,30)[]{$G$}
\Text(80,60)[l]{$: -\frac{\displaystyle 3ig^2}{\displaystyle 4M^2_A}\,
( M^2_H - M^2_G )$}

\DashLine(230,90)(260,60){3}
\DashLine(230,30)(260,60){3}
\DashLine(260,60)(290,90){3}
\DashLine(260,60)(290,30){3}
\Text(225,90)[r]{$H$}
\Text(225,30)[r]{$H$}
\Text(300,90)[]{$G$}
\Text(300,30)[]{$G$}
\Text(310,60)[l]{$: -\frac{\displaystyle ig^2}{\displaystyle 4M^2_A}\,
( M^2_H - M^2_G + 2\xi M^2_A)$}

\Text(220,0)[t]{\bf Fig.\ 2}
\end{picture}
\end{center}

\end{document}